\documentclass[12pt]{article}

\usepackage{amsmath}
\usepackage{amsfonts}
\usepackage{amssymb}
\usepackage{graphicx}
\usepackage{psfrag}
\usepackage{graphicx}
\usepackage{psfrag}

\newcommand{\be}{\begin{equation}}
\newcommand{\ee}{\end{equation}}
\newcommand{\ba}{\begin{eqnarray}}
\newcommand{\ea}{\end{eqnarray}}

\begin{document}
\begin{center}
{\bf
Solvable Two-dimensional Dirac Equation with Matrix Potential: Graphene in External Electromagnetic Field.
}\\
\vspace{1cm}
{\large M. V. Iof\/fe$^{1,}$\footnote{E-mail: ioffe000@gmail.com; corresponding author},
D. N. Nishnianidze}$^{2,}$\footnote{E-mail: cutaisi@yahoo.com}\\
\vspace{0.5cm}
$^1$ Saint Petersburg State University, 7/9 Universitetskaya nab., St.Petersburg, 199034 Russia.\\
$^2$ Akaki Tsereteli State University, 4600 Kutaisi, Georgia.\\

\end{center}
\vspace{0.5cm}
\hspace*{0.5in}
\vspace{1cm}
\hspace*{0.5in}
\begin{minipage}{5.0in}
{\small
It is known that the excitations in graphene-like materials in external electromagnetic field are described by solutions of massless two-dimensional Dirac equation which includes both Hermitian off-diagonal matrix and scalar potentials. Up to now, such two-component wave functions were calculated for different forms of external potentials but, as a rule, depending on one spatial variable only. Here, we shall find analytically the solutions for a wide class of combinations of matrix and scalar external potentials which physically correspond to applied mutually orthogonal magnetic and longitudinal electrostatic fields, both depending really on two spatial variables. The main tool for this progress was provided by supersymmetrical (SUSY) intertwining relations, namely, by their most general - asymmetrical - form proposed recently by the authors. Such SUSY-like method is applied in two steps similarly to the second order factorizable (reducible) SUSY transformations in ordinary Quantum Mechanics.
}
\end{minipage}

Keywords: two-dimensional Dirac equation, SUSY intertwining relations, matrix potential, graphene in electromagnetic field

{\it PACS:} 03.65.-w; 73.22.Pr

\section{Introduction.}

The extensive study of two-dimensional massless Dirac equation in the presence of external electromagnetic fields \cite{efetov-1},
\cite{efetov-2}, \cite{matulis}, \cite{jakubsky-2}, \cite{jakubsky-3}, \cite{jakubsky-1}, \cite{IN} 
 is due to its connection with the properties of electron carriers in graphene and in graphene-like materials \cite{novoselov-2}, \cite{novoselov}, \cite{advances}, \cite{kats}. The actual task is to find analytically the normalizable solutions of such Dirac equation where two components of "spinor" wave function $\Psi (\vec x)\equiv (\Psi_A(\vec x),\,\Psi_B(\vec x))$ correspond to two sublattices of graphene. The potentials in the equation have different origin: the off-diagonal matrix term is provided by the electromagnetic vector-potential $\vec A =(A_1 (x_1, x_2), A_2 (x_1, x_2), 0)$ in the "long derivatives" and leads to the magnetic field $\vec B = \vec\nabla\times\vec A$ along the z-axis, while the scalar potential $A_0 (x_1, x_2)$ describes the interaction with electrostatic or some another scalar field:
\be
[\sigma_1(-i\partial_1-A_1(\vec x)) + \sigma_2(-i\partial_2-A_2(\vec x)) + A_0(\vec x)]\Psi(\vec x)=0,   \label{dirac}
\ee
where two-dimensional $\vec x\equiv (x_1,\, x_2),$ the derivatives $\partial_i\equiv \frac{\partial}{\partial x_i},$ and the charge was taken $e=1.$


It is clear that the case of pure magnetic field in massless two-dimensional Dirac equation is explicitly solvable. Indeed, multiplying (\ref{dirac}) without the term $A_0$ by $\sigma_1,$ one obtains a pair of decoupled first order equations which can be easily solved. The presence in (\ref{dirac}) of term proportional to the unity matrix $\sigma_0$ prevents this decoupling (analogous problem would appear also in presence of the mass term proportional to $\sigma_3$). Different methods were used \cite{peres} - \cite{2-part} to study such two-dimensional Dirac equation, mainly with strong restrictions on the conditions of the problem. There were problems with only electrostatic or only magnetic fields, and also the problems with different specific one-dimensional ansatzes for external fields depending on a variable $x_1,$ or radial variable $r.$ 

The supersymmetrical method inherent in Schr\"odinger Quantum Mechanics \cite{cooper}, \cite{david}, \cite{AI} has become also one of the most effective tools in discussed problems \cite{midya}, \cite{roy}, \cite{ho-1}, \cite{setare}, \cite{schulze2019},\cite{schulze2021},     \cite{kuru2023}. As a rule, the main ingredient of SUSY Quantum Mechanics - so called SUSY intertwining relations - was explored in different forms. In the present paper, a class of external electromagnetic fields will be chosen with both magnetic (matrix) and electrostatic (scalar) terms depend effectively on both spatial coordinates. Such a progress is possible due 
to using a particular form of this approach - so called asymmetric intertwining relations \cite{INP}, \cite{jakubsky2020}, \cite{IN2022}. Up to now, this technique was explored to study the massless two-dimensional Dirac equation with scalar potential and also the equation of Fokker-Planck \cite{fokker}. More specifically, the procedure will include two steps (see Section 2). At first, asymmetric intertwining relations will provide SUSY diagonalization of potential in Eq.(\ref{dirac}) with both electromagnetic and electrostatic two-dimensional terms (Section 3). At the second stage, an additional asymmetric intertwining will connect the Dirac operator with diagonal potential with its partner whose potential is also diagonal but with constant elements (Section 4). 
The solutions of such Dirac equation can be found analytically, and solutions of initial problem are built by applying intertwining operators of both used steps (Section 5). Actually, the whole procedure realizes asymmetric form of factorizable SUSY intertwining of second order. Such SUSY intertwinings, but in their standard (symmetric) form, are known 
\cite{AI} in the context of SUSY Quantum Mechanics with Schr\"odinger operator.

\section{Asymmetric intertwining for two-dimensional Dirac equation in electromagnetic field.}

We start with the asymmetric intertwining relations:
\be 
D_1 N_1=N_2 D_2 \label{int}
\ee
for a pair of two-dimensional massless Dirac operators of the form (\ref{dirac}) rewritten as general operators with Hermitian matrix potentials
\be
D_{1,2} \equiv -i\sigma_k\partial_k + V_{1,2}(\vec x); \quad k=1,2; \quad V_{i}(\vec x) = \left(
                                                                                    \begin{array}{cc}
                                                                                      v^{(i)}_{11}(\vec x) & v^{(i)}_{12}(\vec x) \\
                                                                                      v^{(i)}_{21}(\vec x) & v^{(i)}_{22}(\vec x) \\
                                                                                    \end{array}
                                                                                  \right); \quad i=1,2. 
                                                                                  \label{D}
\ee
Here, two different intertwining operators have the general matrix form:
\be 
N_1 = A_k\partial_k + A(\vec x); \quad N_2 = B_k\partial_k + B(\vec x) \label{NM}
\ee
with constant matrices $A_k, \, B_k$ and two $\vec x-$dependent matrices $A(\vec x), B(\vec x):$
\be 
A(\vec x)=\left(
            \begin{array}{cc}
              a_{11}(\vec x) & a_{12}(\vec x) \\
              a_{21}(\vec x) & a_{22}(\vec x) \\
            \end{array}
          \right); \quad B(\vec x)=\left(
            \begin{array}{cc}
              b_{11}(\vec x) & b_{12}(\vec x) \\
              b_{21}(\vec x) & b_{22}(\vec x) \\
            \end{array}
          \right).
          \label{AABB}
\ee

Expanding Eq.(\ref{int}) in powers of derivatives we obtain:
\ba 
&&\sigma_kA_n-B_k\sigma_n+\sigma_nA_k-B_n\sigma_k=0,\quad k,n=1,2, \label{1-2}\\
&&i\sigma_kA(\vec x)-V_1(\vec x)A_k=-B_kV_2(\vec x)+iB(\vec x)\sigma_k,\quad k=1,2, \label{1-3}\\
&&i\sigma_k(\partial_kA(\vec x))-V_1(\vec x)A(\vec x)=-B_k(\partial_kV_2(\vec x))-B(\vec x)V_2(\vec x),\quad k=1,2. \label{1-4}
\ea
Equations (\ref{1-2}) provide the following form of constant coefficient matrices $A_k,\,B_k$:
\be 
A_1=\left(
      \begin{array}{cc}
        a & b \\
        d & c \\
      \end{array}
    \right); \quad A_2=\left(
      \begin{array}{cc}
        n & -ib \\
        id & p \\
      \end{array}
    \right); \quad B_1=\left(
      \begin{array}{cc}
        c & d \\
        b & a \\
      \end{array}
    \right); \quad B_2=\left(
      \begin{array}{cc}
        p & -id \\
        ib & n \\
      \end{array}
    \right); \label{AB}
\ee
with constant values $a,b,c,d,n,p,$ and Eqs.(\ref{1-3}) give the system of eight linear equations for the matrix elements of $V_{1,2}(\vec x), A(\vec x), B(\vec x):$
\ba  
&&(p+ic)(v^{(1)}_{12}-v^{(2)}_{12})=0; \label{3-2}\\
&&(n-ia)(v^{(1)}_{21}-v^{(2)}_{21})=0; \label{4-3}\\
&&(n+ia)v^{(1)}_{11}+2idv^{(1)}_{12}-(p+ic)v^{(2)}_{11}=2b_{12}; \label{3-1}\\
&&(n+ia)(v^{(1)}_{21}-v^{(2)}_{21})+2idv^{(1)}_{22}-2ibv^{(2)}_{11}=-2a_{11}+2b_{22}; \label{3-3}\\
&&(p+ic)v^{(1)}_{22}-2ibv^{(2)}_{12}-(n+ia)v^{(2)}_{22}=-2a_{12}; \label{3-4}\\
&&(n-ia)v^{(1)}_{11}+2idv^{(2)}_{21}-(p-ic)v^{(2)}_{11}=2a_{21}; \label{4-1}\\
&&(p-ic)(v^{(1)}_{12}-v^{(2)}_{12})+2idv^{(2)}_{22}-2ibv^{(1)}_{11}=-2b_{11}+2a_{22}; \label{4-2}\\
&&(p-ic)v^{(1)}_{22}-(n-ia)v^{(2)}_{22}-2ibv^{(1)}_{21}=-2b_{21}.  \label{4-4}
\ea 
The first two equations of the system indicates the need to highlight four different possibilities for solution of (\ref{3-2}) -- (\ref{4-4}) with Hermiticity of both potentials in mind:
\ba 
I) &&p+ic=n-ia=0; \nonumber\\
II) &&v^{(1)}_{12}=v^{(2)}_{12}; \quad v^{(1)}_{21}=v^{(2)}_{21}; \quad (p+ic)(n-ia) \neq 0\nonumber\\
III) &&v^{(1)}_{12}=v^{(2)}_{12}; \quad v^{(1)}_{21}=v^{(2)}_{21};\quad p+ic=0; \nonumber\\
IV) &&v^{(1)}_{12}=v^{(2)}_{12}; \quad v^{(1)}_{21}=v^{(2)}_{21};\quad n-ia=0. \nonumber
\ea 

Thus, Eqs.(\ref{1-2}), (\ref{1-3}) are reduced to the system of Eqs.(\ref{3-1}) - (\ref{4-4}) in four possible variants above. We have not considered yet the matrix differential equation (\ref{1-4}). It is convenient to solve it in an indirect form by using its combination with derivative $\partial_k$ of equation (\ref{1-3}):
\be 
V_1(\vec x)A(\vec x) - B(\vec x)V_2(\vec x)=(\partial_kV_1(\vec x))A_k+i(\partial_kB(\vec x))\sigma_k.
\label{1-9}
\ee

Below, all four variants will be used to obtain the general solution of initial intertwining relations (\ref{int}).

\section{SUSY-diagonalization by means of intertwining.}

Substitution of $p+ic=n-ia=0,$ i.e. of Variant I above, into other equations (\ref{3-1}) - (\ref{4-4}) together with Eqs.(\ref{1-9}) is quite long, but straightforward. These calculations lead to a system of four nonlinear equations, two of them being differential:
\ba 
v^{(1)}_{11}(ia_{11}+bv^{(2)}_{11}+av^{(2)}_{21})&=&v^{(2)}_{11}(ia_{22}+cv^{(2)}_{12}+dv^{(2)}_{22}); \label{10-1-1}\\
v^{(1)}_{22}(ia_{22}+cv^{(2)}_{12}+dv^{(2)}_{22})&=&v^{(2)}_{22}(ia_{11}+bv^{(2)}_{11}+av^{(2)}_{21}); \label{10-1-2}\\
i(v^{(2)}_{12}-v^{(1)}_{12}) (ia_{22}+cv^{(2)}_{12}+dv^{(2)}_{22})&=&2\partial_z(ia_{22}+cv^{(2)}_{12}+dv^{(2)}_{22}); \label{10-2-1}\\
i(v^{(2)}_{21}-v^{(1)}_{21}) (ia_{11}+bv^{(2)}_{11}+av^{(2)}_{21})&=&2\partial_{\bar z}(ia_{11}+bv^{(2)}_{11}+av^{(2)}_{21}). \label{10-2-2}
\ea 
Here and below, it is convenient to use the space arguments $\vec x$ of the functions in the form of complex variables $z=x_1+ix_2; \quad \bar z=x_1-ix_2$ and corresponding derivatives $\partial\equiv \frac{1}{2}(\partial_1-i\partial_2);\,\,\bar\partial\equiv \frac{1}{2}(\partial_1+i\partial_2)$. In particular, the system (\ref{10-1-1}) - (\ref{10-2-2}) takes the compact form:
\ba 
 v^{(1)}_{11}(z, \bar z) f_1(z, \bar z) &=& v^{(2)}_{11}(z, \bar z) f_2(z, \bar z); \label{11-1}\\
v^{(1)}_{22}(z, \bar z) f_2(z, \bar z) &=& v^{(2)}_{22}(z, \bar z) f_1(z, \bar z); \label{11-2}\\
v^{(1)}_{12}(z, \bar z)&=&v^{(2)}_{12}(z, \bar z)+2i\partial\ln (f_2(z,\bar z)); \label{11-3}\\
v^{(1)}_{21}(z, \bar z)&=&v^{(2)}_{21}(z, \bar z)+2i\bar\partial\ln (f_1(z,\bar z)), \label{11-4}
\ea 
where two combinations are introduced:
\ba 
f_1(z, \bar z)\equiv ia_{11}(z, \bar z)+bv^{(2)}_{11}(z,\bar z)+av^{(2)}_{21}(z, \bar z); \label{11-0}\\
f_2(z, \bar z)\equiv ia_{22}(z, \bar z)+cv^{(2)}_{12}(z,\bar z)+dv^{(2)}_{22}(z, \bar z). \label{11-0-0}
\ea 

One may notice the apparent paradox contained in the system (\ref{10-1-1}) - (\ref{10-2-2}). All these equations are identically fulfilled if
$$ia_{11}+bv^{(2)}_{11}+av^{(2)}_{21}=ia_{22}+cv^{(2)}_{12}+dv^{(2)}_{22}=0$$
for arbitrary potentials $V_1,\, V_2.$ The explanation is rather simple: just in this case, Dirac operators $D_1, D_2$ are proportional to the intertwining operators $D_2=CN_1, D_1=N_2C$ up to some constant matrix $C.$ Therefore, intertwining relation (\ref{int}) becomes a trivial identity in such a case.

In the present context, the typical way \cite{cooper}, \cite{david}, \cite{AI}, \cite{INP}, \cite{IN2022} of using SUSY intertwining relations can be formulated as follows. Let us choose one of the Dirac operators, $D_2,$ such that the corresponding Dirac equation is rather simple so that the problem of its analytical solution is more easy. Then, the solutions of partner Dirac equation with operator $D_1$ will be found by the action of intertwining operator $\Psi^{(1)}=N_1\Psi^{(2)}$ (in its turn, the operator $N_2^{\dagger}$ transforms the spinor $\Psi^{(1)}$ into $\Psi^{(2)}$). As the first step on this way, we choose the potential $V_2(\vec x)$ as a diagonal matrix:
\be 
V_2(\vec x) = diag(v^{(2)}_{11}(z,\bar z),\,\, v^{(2)}_{22}(z,\bar z))\equiv diag(v_1(z,\bar z),\,\, v_2(z,\bar z))
\label{V_2}
\ee
with real diagonal matrix elements. Then, due to Eqs.(\ref{11-1}), (\ref{11-2}), fraction $f_2(z, \bar z)/f_1(z, \bar z)$ is a real function. Futhermore,  Eqs.(\ref{11-3}), (\ref{11-4}) and the Hermiticity of $V_1(\vec x),$ i.e. $v^{(1)}_{12}(z,\bar z)=v^{(1)\star}_{21}(z,\bar z),$ lead to the following restriction:
\be 
f_1(z, \bar z)f^{\star}_2(z, \bar z)=c
\label{12-2}
\ee
with an arbitrary real constant $c.$ Summarizing these results, potential $V_1$ is expressed in terms of components (\ref{V_2}), function $f_2(z, \bar z)$ and real constant $c:$
\be 
V_1(\vec x)=\left(
              \begin{array}{cc}
                \frac{|f_2|^2}{c}v_1 &  2i\partial\ln (f_2)\\ 
                -2i\bar\partial\ln (f_2^{\star}) & \frac{c}{|f_2|^2}v_2 \\
              \end{array}
            \right)= \left(
              \begin{array}{cc}
                \pm f^2v_1 &  2i(\partial\ln (f) +i\partial\varphi)\\ 
                -2i(\bar\partial\ln (f) -i\bar\partial\varphi )& \pm f^{-2}v_2 \\
              \end{array}
            \right),
\label{12-4}
\ee
where the function $f_2$ was parameterizing as:
\be 
f_2(z, \bar z)\equiv \sqrt{\pm c}f(z, \bar z)e^{i\varphi(z, \bar z)}. \label{12-12}
\ee
Above, the sign $\pm$ corresponds to cases $c > 0,\, c < 0,$ respectively, and $f(z, \bar z)$ is a positive function. This function 
connects elements of initial diagonal potential $V_2$ as:
\be 
v_2(z, \bar z) = f^4(z, \bar z)v_1(z, \bar z).
\label{13-2}
\ee

For the physical system with matrix potential $V_1$ describing "spin $1/2$ particle" in external electromagnetic field according to Eq.(\ref{dirac}), the diagonal elements of $V_1$ define the electrostatic potential
\be 
A_0(\vec x)=v^{(1)}_{11}=v^{(1)}_{22}=\pm (v_1 v_2)^{1/2}=\pm f^2(\vec x)v_1(\vec x),
\label{A_0}
\ee
while the off-diagonal terms define the magnetic field which is orthogonal to the plane $x_1,x_2$:
\be 
B_3(\vec x)=\vec\nabla\times\vec A(\vec x)=\bigtriangleup \ln(f(\vec x)).
\label{13-6}
\ee
The solutions of Dirac equation with such external fields could be obtained from the two-component solutions $\Psi^{(2)}(\vec x)$ of Dirac equation with diagonal potential $V_2$ by action of the intertwining operator $N.$ But the diagonal form of potential $V_2$ does not yet provide solvability of the corresponding Dirac problem analytically. Up to now, supersymmetric intertwining relations (\ref{int}) connected an initial Dirac problem with operator $D_1$ to Dirac problem with the potential $V_2,$ which is diagonal, and therefore, has chances to be solvable. This part of procedure can be called SUSY-diagonalization of two-dimensional Dirac problem with matrix potential (by analogy with SUSY separation of variables \cite{new}).

 \section{From diagonal potential to the constant one by means of intertwining.}
 
 Let us consider now the variant II with $(p+ic)(n-ia)\neq 0$ for solution of the system (\ref{3-2})  - (\ref{4-4}) in the context of the second step of our procedure. Namely, let us consider intertwining relations between two Dirac operators, both with diagonal potential:
 \be 
U_1(\vec x)=
  \left(
              \begin{array}{cc}
                v_{1}(\vec x) & 0\\ 
                0 & v_{2}(\vec x) \\
              \end{array}
            \right); 
            \quad U_2(\vec x)= \left(
              \begin{array}{cc}
                m_1 & 0\\ 
                0 & m_2 \\
              \end{array}
            \right),
\label{01-0}
\ee
with constant elements $m_1, m_2.$ Here, potential $U_1(\vec x)$ is identified with potential $V_2(\vec x)$ of the previous step, and the constant partner potential $U_2$ will provide solvability of the problem. The intertwining operators $N_1, \, N_2$ have the same general form (\ref{NM}), and the explicit expressions for matrices $A_k,\, B_k,\, A(\vec x),\, B(\vec x)$ will be found below by analytical solution of the system of equations which were obtained in Section 2. 

We shall consider the system of equations (\ref{3-2}) - (\ref{4-4}) sequentially. Eqs.(\ref{3-2}) and (\ref{4-3}) are fulfilled automatically. Eqs.(\ref{3-1}), (\ref{4-1}) and (\ref{3-4}), (\ref{4-4}) allow to express off-diagonal elements of $A(\vec x),\, B(\vec x)$ at (\ref{AABB}) in terms of $v_1(\vec x), \, v_2(\vec x):$
\ba 
a_{12}(\vec x)&=&-\frac{1}{2}(p+ic)v_2(\vec x) + \frac{1}{2}m_2(n+ia); \label{03-2}\\
a_{21}(\vec x)&=&\frac{1}{2}(n-ia)v_2(\vec x) - \frac{1}{2}m_1(p-ic); \label{03-1}\\
b_{12}(\vec x)&=&\frac{1}{2}(n+ia)v_1(\vec x) - \frac{1}{2}m_1(p+ic); \label{03-3}\\
a_{12}(\vec x)&=&-\frac{1}{2}(p-ic)v_2(\vec x) + \frac{1}{2}m_2(n-ia). \label{03-4}
\ea
Eqs.(\ref{4-2}) and (\ref{3-3}) for diagonal elements of $A(\vec x),\, B(\vec x)$ can be written as:
\ba 
b_{11}(\vec x)=a_{22}(\vec x)+ibv_1(\vec x)-idm_2; \label{04-5}\\
b_{22}(\vec x)=a_{11}(\vec x)+idv_2(\vec x)-ibm_1. \label{04-6}
\ea

Eq.(\ref{1-4}) in its initial form is convenient to write now in complex coordinates $z\, \bar z:$
\ba  
&&2i\Biggl[\left(
              \begin{array}{cc}
                0 & \partial\\ 
                \bar{\partial} & 0 \\
              \end{array}
            \right) - \left(
              \begin{array}{cc}
                v_{1}(z,\bar z) & 0\\ 
                0 & v_{2}(z,\bar z) \\
              \end{array}
            \right)\Biggr] \left(
              \begin{array}{cc}
                a_{11}(z,\bar z) & a_{12}(z,\bar z)\\ 
                a_{21}(z,\bar z) & a_{22}(z,\bar z) \\
              \end{array}
            \right) = \nonumber \\ 
            &&=-\left(
              \begin{array}{cc}
                b_{11}(z,\bar z) & b_{12}(z,\bar z)\\ 
                b_{21}(z,\bar z) & b_{22}(z,\bar z) \\
              \end{array}
            \right) \left(
              \begin{array}{cc}
                m_1 & 0\\ 
                0 & m_2 \\
              \end{array}
            \right),
\label{04-0}
\ea 
where $\partial\equiv\partial_z=\frac{1}{2}(\partial_1-i\partial_2),\,\,\bar{\partial}\equiv\partial_{\bar z}=\frac{1}{2}(\partial_1+i\partial_2).$
In components, matrix equation (\ref{04-0}) is equivalent to the system of linear first order differential equations:
\ba 
2i(\partial a_{21}(z, \bar z))-v_1(z, \bar z)a_{11}(z, \bar z)&=&-m_1b_{11}(z, \bar z); 
\label{04-1}\\
2i(\bar{\partial} a_{12}(z, \bar z))-v_2(z, \bar z)a_{22}(z, \bar z)&=&-m_2 b_{22}(z, \bar z); 
\label{04-2}\\
2i(\partial a_{22}(z, \bar z))-v_1(z, \bar z)a_{12}(z, \bar z)&=&-m_2b_{12}(z, \bar z); 
\label{04-3}\\
2i(\bar{\partial} a_{11}(z, \bar z))-v_2(z, \bar z)a_{21}(z, \bar z)&=&-m_1b_{21}(z, \bar z). 
\label{04-4}
\ea
Let us define for convenience:
\be 
a_{11}(z, \bar z)-im_1b\equiv g_1(z, \bar z); \quad a_{22}(z, \bar z)-im_2d\equiv g_2(z, \bar z).
\label{06-0}
\ee
Then, after substitution of Eqs.(\ref{03-2}) - (\ref{03-4}) and Eqs.(\ref{04-5}), (\ref{04-6}), the system (\ref{04-1}) - (\ref{04-4}) takes the form:
\ba 
&&i(n-ia)(\partial v_{1}(z, \bar z))=v_1(z, \bar z)g_{1}(z, \bar z)-m_1g_{2}(z, \bar z); \label{06-1}\\
&&i(p+ic)(\bar{\partial} v_{2}(z, \bar z))=-v_2(z, \bar z)g_{2}(z, \bar z)+m_2g_{1}(z, \bar z); \label{06-2}\\
&&4i(\bar\partial g_{1}(z, \bar z))=(n-ia)(v_1(z, \bar z)v_2(z, \bar z)-m_1m_2); \label{06-3}\\
&&4i(\partial g_{2}(z, \bar z))=-(p+ic)(v_1(z, \bar z)v_2(z, \bar z)-m_1m_2), \label{06-4}
\ea
where the form of last two equations means that functions $g_1(z, \bar z),\, g_2(z, \bar z)$ can be expressed in terms of one complex function $g$:
\be 
g_1(z, \bar z)=-\frac{1}{p+ic}(\partial g(z, \bar z));\quad g_2(z, \bar z)=\frac{1}{n-ia}(\bar{\partial} g(z, \bar z)),
\label{06-34}
\ee
which satisfies the second order equation:
\be 
4i(\partial\bar{\partial}g(z, \bar z))=-(n-ia)(p+ic)(v_1(z, \bar z) v_2(z, \bar z)-m_1m_2).
\label{06-5}
\ee
and the first two equations (\ref{06-1}), (\ref{06-2}) become now:
\ba  
&&i(n-ia)(\partial v_{1}(z, \bar z))=-\frac{1}{p+ic}v_1(z, \bar z) (\partial g(z, \bar z))-\frac{1}{n-ia}m_1(\bar{\partial}g(z, \bar z)); \label{06-6}\\
&&i(p+ic)(\bar{\partial} v_{2}(z, \bar z))=-\frac{1}{n-ia}v_2(z, \bar z) (\bar{\partial} g(z, \bar z))-\frac{1}{p+ic}m_2(\partial g(z, \bar z)). \label{06-7}
\ea
Thus,
\ba 
&&\Omega (\partial v_1(z, \bar z)) + v_1(z, \bar z)(\partial g(z, \bar z)) = -L_1 (\bar{\partial} g(z, \bar z));
\label{07-2}\\
&&\Omega (\bar{\partial} v_2(z, \bar z)) + v_2(z, \bar z)(\bar{\partial} g(z, \bar z)) = -L_2 (\partial g(z, \bar z));
\label{07-3},\\
&&\Omega\equiv i(n-ia)(p+ic);\quad L_1\equiv\frac{m_1(p+ic)}{n-ia};\quad L_2\equiv\frac{m_2(n-ia)}{p+ic}\nonumber
\ea
Below, we shall solve this system of equation by considering separately different options for the choice of constant parameters.

\subsection{{\bf The case A:}  Parameters are real.}
 
Let us study the case with real function $g(z, \bar z)$ and real values of parameters $\Omega,\, L_1, \, L_2.$ Taking into account the reality 
of $v_1(z, \bar z)$ and $v_2(z, \bar z),$ it is useful to come back to Cartesian coordinates $x_1,\,x_2$ in Eqs.(\ref{07-2}) and (\ref{07-3}). Separately, both real and imaginary parts of (\ref{07-2}) are integrated explicitly with two "constants of integration" $s_1(x_1),\, s_2(x_2),$ which are arbitrary real functions of their arguments:
\ba 
\exp{(g(\vec x)/\Omega)} &=& L_1^{-1}(s_2(x_2)-s_1(x_1)); \label{08-11}\\
v_1(\vec x)&=&L_1\frac{s_2(x_2)+s_1(x_1)}{s_2(x_2)-s_1(x_1)}. \label{08-12}
\ea
Analogously, Eq.(\ref{07-3}) can be integrated as well with similar result:
\ba 
\exp{(g(\vec x)/\Omega)} &=& L_2^{-1}(\tilde{s}_2(x_2)-\tilde{s}_1(x_1)); \label{08-21}\\
v_2(\vec x)&=&L_2\frac{\tilde{s}_2(x_2)+\tilde{s}_1(x_1)}{\tilde{s}_2(x_2)-\tilde{s}_1(x_1)}, \label{08-22}
\ea
and arbitrary real $\tilde{s}_1(x_1), \,\tilde{s}_2(x_2).$
Eqs.(\ref{08-11}), (\ref{08-21}) together allow to connect $\tilde{s}_1(x_1), \,\tilde{s}_2(x_2)$ with their analogues:
\be 
\tilde{s}_1(x_1)=L_1^{-1}L_2(s_1(x_1)+\delta);\quad \tilde{s}_2(x_2)=L_1^{-1}L_2(s_2(x_2)+\delta),
\label{08-3}
\ee
with an arbitrary real constant $\delta .$ These relations have to be substituted into expression (\ref{08-22}).

Using these connections in second order differential equation (\ref{06-5}) for function $g(\vec x)$ and differentiating it by $\partial_1\partial_2,$ we obtain the simple third order equation with separable variables:
\be 
\frac{s_1^{\prime\prime\prime}(x_1)}{s_1^{\prime}(x_1)} + \frac{s_2^{\prime\prime\prime}(x_2)}{s_2^{\prime}(x_2)} = -4L_1L_2=-4m_1m_2.
\label{09-3}
\ee
After separation of variables in (\ref{09-3}) and integration of one-dimensional equations, we have:
\be 
s_1^{\prime\prime}(x_1)=\lambda_1^2s_1(x_1)+\omega_1\lambda_1^2; \quad   
s_2^{\prime\prime}(x_2)=\lambda_2^2s_2(x_2)+\omega_2\lambda_2^2,
\label{09-4}
\ee
where $\omega_1,\,\omega_2$ are integration constants, and $\lambda_1,\,\lambda_2$ - arbitrary constants which satisfy relation:
\be 
\lambda_1^2 + \lambda_2^2=-4m_1m_2.\nonumber 
\ee
Solutions of (\ref{09-4}) are known:
\ba 
s_1(x_1)&=&\frac{1}{2}(\sigma_1e^{\lambda_1x_1}+\delta_1e^{-\lambda_1x_1}) - \omega_1; \nonumber \\
s_2(x_2)&=&\frac{1}{2}(\sigma_2e^{\lambda_2x_2}+\delta_2e^{-\lambda_2x_2}) - \omega_2.\nonumber 
\ea
Because we used derivatives of Eq.(\ref{06-5}), it is necessary to check results. Substitution of (\ref{09-4}) into (\ref{06-5}) gives two relations between parameters:
\ba 
&&\omega_1\lambda_1^2 + \omega_2\lambda_2^2 + 2 \delta L_1L_2 = 0; \nonumber \\
&&\lambda_1^2(\sigma_1\delta_1-\omega_1^2) +\lambda_2^2(\sigma_2\delta_2-\omega_2^2)=0. \nonumber 
\ea

Now we can list, depending on the values of the constants, all possible solutions $s_1(x_1),\, s_2(x_2)$ within this Subsection. All of them are expressed in terms of hyperbolic, trigonometric and exponential functions, and they have to be inserted into (\ref{08-3}) to find $u_1(\vec x),\, u_2(\vec x)$ according to (\ref{08-12}), (\ref{08-22}):

{\bf I.}  $\lambda_1^2>0;\quad \lambda_2^2>0; \quad \sigma_1\delta_1>0; \quad \sigma_2\delta_2>0$
  
  By additional translation of $x_{1,2}$ functions $s_1(x_1),\, s_2(x_2)$ takes the form:
  \be 
  s_1(x_1)=\sigma_1\cosh(\lambda_1x_1)-\omega_1;\quad s_2(x_2)=\sigma_2\cosh(\lambda_2x_2)-\omega_2, \nonumber 
  \ee
  with restriction:
  \be 
  \lambda_1^2(\sigma_1^2-\omega_1^2) + \lambda_2^2(\sigma_2^2-\omega_2^2)=0. \nonumber 
  \ee
  
{\bf II.}  $\lambda_1^2>0;\quad \lambda_2^2>0; \quad \sigma_1\delta_1>0; \quad \sigma_2\delta_2<0$
  \be 
  s_1(x_1)=\sigma_1\cosh(\lambda_1x_1)-\omega_1;\quad s_2(x_2)=\sigma_2\sinh(\lambda_2x_2)-\omega_2, \nonumber 
  \ee
  with restriction:
  \be 
  \lambda_1^2(\sigma_1^2-\omega_1^2) - \lambda_2^2(\sigma_2^2+\omega_2^2)=0.\nonumber 
  \ee
{\bf III.}  $\lambda_1^2>0;\quad \lambda_2^2>0; \quad \sigma_1\delta_1>0; \quad \delta_2=0$
  \be 
  s_1(x_1)=\sigma_1\cosh(\lambda_1x_1)-\omega_1;\quad s_2(x_2)=\frac{1}{2}\sigma_2\exp(\lambda_2x_2)-\omega_2, \nonumber 
  \ee
  with restriction:
  \be 
  \lambda_1^2(\sigma_1^2-\omega_1^2) - \lambda_2^2\omega_2^2=0. \nonumber 
  \ee

{\bf IV.}  $\lambda_1^2<0;\quad \lambda_2^2<0; \quad \sigma_1=\delta_1^{\star}; \quad \sigma_2=\delta_2^{\star}$
 
 $\lambda_1\equiv i\Lambda_1;\quad \lambda_2\equiv i\Lambda_2$
  \be 
  s_1(x_1)=\sigma_1\cos(\Lambda_1x_1)-\omega_1;\quad s_2(x_2)=\sigma_2\cos(\Lambda_2x_2)-\omega_2; \nonumber 
  \ee
  with restriction:
  \be 
  \Lambda_1^2(\sigma_1^2-\omega_1^2) + \Lambda_2^2(\sigma_2^2-\omega_2^2)=0.\nonumber 
  \ee

{\bf V.}  $\lambda_1^2>0;\quad \lambda_2^2<0; \quad \sigma_1\delta_1>0; \quad \sigma_2=\delta_2^{\star}$
 
 $\lambda_2\equiv i\Lambda_2$
  \be 
  s_1(x_1)=\sigma_1\cosh(\lambda_1x_1)-\omega_1;\quad s_2(x_2)=\sigma_2\cos(\Lambda_2x_2)-\omega_2; \nonumber 
  \ee
  with restriction:
  \be 
  \lambda_1^2(\sigma_1^2-\omega_1^2) - \Lambda_2^2(\sigma_2^2-\omega_2^2)=0. \nonumber 
  \ee
  
  {\bf VI.}  $\lambda_1^2>0;\quad \lambda_2^2<0; \quad \sigma_1\delta_1<0; \quad \sigma_2=\delta_2^{\star}$
 $\lambda_2\equiv i\Lambda_2$
  \be 
  s_1(x_1)=\sigma_1\sinh(\lambda_1x_1)-\omega_1;\quad s_2(x_2)=\sigma_2\cos(\Lambda_2x_2)-\omega_2; \nonumber 
  \ee
  with restriction:
  \be 
  \lambda_1^2(\sigma_1^2+\omega_1^2) + \Lambda_2^2(\sigma_2^2-\omega_2^2)=0. \nonumber 
  \ee
  
\subsection{{\bf The case B:} $m_2=0.$}

The case with $m_2=L_2=0$ and again real $\Omega,\, L_1, \, g(\vec x)$ will be considered below. For such a choice, Eq.(\ref{07-3}) gives:
\be 
v_2(\vec x)=\exp(g(\vec x)/\Omega)=\frac{C}{s_2(x_2)-s_1(x_1)}, \quad v_1(\vec x)=L_1\frac{s_1(x_1)+s_2(x_2)}{s_2(x_2)-s_1(x_1)},\label{014-3}
\ee
and Eq.(\ref{06-5}) looks like:
\be 
(s_2^{\prime\prime}(x_2)-s_1^{\prime\prime}(x_1))(s_2(x_2)-s_1(x_1))-((s_1^{\prime}(\vec x_1))^2+(s_2^{\prime}(\vec x_2))^2)=CL_1(s_1(x_1)+s_2(x_2)).
\label{014-4}
\ee
The latter equation after differentiation is again, similarly to (\ref{09-3}), amenable to separation of variables but with zero in the r.h.s. It has two different solutions depending on the value of separation constant:
\be
s_1(x_1)=a_1x_1^2+b_1x_1+c_1; \quad s_2(x_2)=a_2x_2^2+b_2x_2+c_2.
 \label{014-6}
\ee
and
\be 
s_1^{\prime\prime}(x_1)=\lambda^2(s_1(x_1)+\omega_1);\quad s_2^{\prime\prime}(x_2)=-\lambda^2(s_2(x_2)+\omega_2) \label{014-5}
\ee
The latter one has three kinds of explicit solutions to insert into (\ref{014-3}):
\ba 
s_1(x_1)&=&\sigma_1\cosh(\lambda x_1)-\omega_1;\quad s_2(x_2)=\sigma_2\cos(\lambda x_2)-\omega_2; \nonumber \\
s_1(x_1)&=&\sigma_1\sinh(\lambda x_1)-\omega_1;\quad s_2(x_2)=\sigma_2\cos(\lambda x_2)-\omega_2; \nonumber \\
s_1(x_1)&=&\frac{1}{2}\sigma_1\exp(\lambda x_1)-\omega_1;\quad s_2(x_2)=\sigma_2\cos(\lambda x_2)-\omega_2; \nonumber 
\ea
with restrictions for constants, correspondingly:
\ba 
&&\omega_2^2-\omega_1^2+\sigma_1^2-\sigma_2^2=0; \nonumber\\
&&\omega_2^2-\omega_1^2-\sigma_1^2-\sigma_2^2=0; \nonumber \\
&&\omega_2^2-\omega_1^2-\sigma_2^2=0; \nonumber\\
&&\lambda^2(\omega_2-\omega_1)-CL_1=0.\nonumber
\ea
As for the polynomial solution (\ref{014-6}), a few restrictions have to be fulfilled simultaneously:
\be 
A\equiv2a_1+2a_2+CL_1=0. \label{016-3}
\ee
Finally, solution (\ref{014-6}) leads to two different opportunities for the components of $U_1(\vec x).$ The first one:
\ba 
v_1(\vec x)&=&-L_1\frac{[CL_1(c_2x_1^2-c_1x_2^2)+2(c_1^2-c_2^2)]}{{CL_1(c_2x_1^2+c_1x_2^2)+2(c_1-c_2)^2}};\label{017-4}\\
v_2(\vec x)&=&-\frac{2(c_1-c_2)}{{CL_1(c_2x_1^2+c_1x_2^2)+2(c_1-c_2)^2}};\label{017-5}\\
a_1&=&\frac{Cc_2L_1}{2(c_1-c_2)}; \quad a_2=-\frac{Cc_1L_1}{2(c_1-c_2)}; \, c_1\neq c_2,\nonumber
\ea
and the second, with $c_1=c_2=0$ and functions $s_1(x_1)=a_1x_1^2,\, s_2(x_2)=-(a_1+\frac{1}{2}CL_1)x_2^2,$ is:
\be 
v_1(\vec x)=-L_1\frac{a_1x_1^2-(a_1+\frac{CL_1}{2})x_2^2}{(a_1+\frac{CL_1}{2})x_2^2+a_1x_1^2};\quad 
v_2(\vec x)=-\frac{C}{(a_1+\frac{CL_1}{2})x_2^2+a_1x_1^2}.
\label{018-1}
\ee

\subsection{{\bf The case C:} $p+ic=0.$}

Let us consider again intertwining of two Dirac operators both with diagonal potential:
\be 
W_1(\vec x)=\left(
              \begin{array}{cc}
                w_{11}^{(1)}(\vec x) & 0\\ 
                0 & w_{22}^{(1)}(\vec x) \\
              \end{array}
            \right) \equiv \left(
              \begin{array}{cc}
                v_{1}(\vec x) & 0\\ 
                0 & v_{2}(\vec x) \\
              \end{array}
            \right); 
            \quad W_2(\vec x)= \left(
              \begin{array}{cc}
                k_1 & 0\\ 
                0 & k_2 \\
              \end{array}
            \right),
\label{001-0}
\ee
with constant elements $k_1, k_2.$ Here, matrix potential $W_1(\vec x)$ will be also identified with potential $V_2(\vec x)$ of Section 3, and the constant partner potential $W_2$ will provide solvability of the problem. This means that the case C is an alternative option in relation to cases B and C of the previous two Subsections. The difference with them is that we will take now $p+ic=0,$ i.e. variant III of Section 2 (it is clear that the variant IV can be considered analogously). In this case, the system (\ref{3-1}) - (\ref{4-4}) is expressed as:
\ba 
b_{11}&=&a_{22}+ibv_1-idk_2;\quad  b_{12}=\frac{n+ia}{2}v_1;\quad b_{21}=-pw_2+\frac{n-ia}{2}k_2; \nonumber \\
a_{12}&=&\frac{n+ia}{2}k_2=Const;\quad a_{21}=-pk_1+\frac{n-ia}{2}v_1. \nonumber 
\ea 
Matrix differential equation (\ref{1-9}) takes a form:
\ba 
i(n-ia)\partial v_1(\vec x)&=&v_1(\vec x)(a_{11}(\vec x)-ibk_1)-k_1(a_{22}(\vec x)-idk_2);\nonumber \\
v_2(\vec x)(a_{22}(\vec x)-idk_2)&=&k_2(a_{11}(\vec x)-ibk_1);\nonumber \\
\partial a_{22}(\vec x)&=&0;\nonumber \\
2i\bar\partial a_{11}(\vec x)&=&\frac{n-ia}{2}(v_1(\vec x)v_2(\vec x)-k_1k_2),  \nonumber
\ea 
which by convenient definition,
\be 
r_1(\vec x)\equiv a_{11}(\vec x)-ibk_1;\quad r_2(\vec x)\equiv a_{22}(\vec x)-idk_2,  \nonumber
\ee 
can be reduced to two equations $(r_2(\vec x)=r_2(\bar z))$:
\ba 
&&ik_2(n-ia)\partial (\frac{v_1(\vec x)}{r_2(\bar z)})=v_1(\vec x)v_2(\vec x)-k_1k_2; \label{003-1}\\
&&4i(\bar\partial r_1(\vec x))=(n-ia)(v_1(\vec x)v_2(\vec x)-k_1k_2). \label{003-2}
\ea 
so that both $v_1(\vec x)$ and $v_2(\vec x)$ are expressed in terms of one function $\kappa (\vec x):$
\be 
v_1(\vec x)=\frac{4r_2(\bar z)}{k_2(n-ia)}(\bar\partial \kappa (\vec x));\quad 
v_2(\vec x)=\frac{k_2(n-ia)}{r_2(\bar z)}(\partial \kappa (\vec x));\quad r_1(\vec x)=(n-ia)(\partial\kappa (\vec x)), \label{003-3}
\ee 
where function $\kappa (\vec x)$ satisfies the following second order differential equation:
\be 
i(\partial\bar\partial\kappa (\vec x))=(\partial\kappa (\vec x))(\bar\partial\kappa (\vec x))-k;\quad 4k\equiv k_1k_2. \label{003-5}
\ee
Since the r.h.s. here is real, the real part of the function $\kappa (\vec x)$ is a sum of two mutually conjugated functions:
\be 
\kappa (\vec x) \equiv \alpha (z) + \bar{\alpha} (\bar z) + i\xi (\vec x). \nonumber 
\ee 
Due to Eq.(\ref{003-3}), the reality of both diagonal elements $v_1(\vec x), \, v_2(\vec x)$ leads to reality 
of $(\partial\kappa(z,\bar z))(\bar\partial\kappa(z,\bar z))$ that means in terms of $\alpha$ and $\xi:$
\be 
\textit{Im}((\partial\kappa (\vec x))(\bar\partial\kappa (\vec x)))=\alpha^{\prime}(z)\bar\partial\xi (\vec x) + \bar\alpha^{\prime}(\bar z)\partial\xi (\vec x)=0, \nonumber 
\ee 
i.e. function $\xi (\vec x)$ is an arbitrary real function of the specific real argument:
\be 
\xi (\vec x)=\Phi (i(\alpha (z)-\bar\alpha (\bar z)))\equiv \Phi (X); \quad X(z, \bar z)\equiv i(\alpha (z)-\bar\alpha (\bar z)). \label{004-2}
\ee 
From Eq.(\ref{003-5}), we obtain nonlinear differential equation for the function $\Phi :$
\be 
\Phi^{\prime\prime}(X)-(\Phi^{\prime}(X))^2+1=\frac{k}{\alpha^{\prime}(z)\bar\alpha^{\prime}(\bar z)}. \label{004-4}
\ee
The l.h.s. in (\ref{004-4}) depends only on the variable $X$ which satisfies the following equation by definition:
\be 
(\frac{1}{\alpha^{\prime}(z)}\partial + \frac{1}{\bar\alpha^{\prime}(\bar z)}\bar\partial )X(z, \bar z)=0, \nonumber
\ee
and therefore, due to (\ref{004-4}):
\be 
(\frac{1}{\alpha^{\prime}(z)}\partial + \frac{1}{\bar\alpha^{\prime}(\bar z)}\bar\partial )\frac{1}{\alpha^{\prime}(z)\bar\alpha^{\prime}(\bar z)}=0,\nonumber 
\ee
allowing to define possible forms of the function $\alpha(z)$. Indeed, variables in the latter equation can be separated:
\be 
\frac{\alpha^{\prime\prime}(z)}{(\alpha^{\prime}(z))^2} + \frac{\bar\alpha^{\prime\prime}(\bar z)}{(\bar\alpha^{\prime}(\bar z))^2}=0 \nonumber
\ee 
informing us that exactly two options exist for the function $\alpha (z):$ 
\be 
a) \,\,\, \alpha^{\prime}(z)=\omega;\quad b)\,\,\, \alpha^{\prime}(z)=\frac{i\lambda}{z},
\label{005-00}
\ee
with $\omega $ - an arbitrary constant and $\lambda $ - an arbitrary real constant.

For the first option, $r_2(\bar z)$ must be constant, 
\ba 
v_1(\vec x)&=&4e(1+\Phi^{\prime}(X(z,\bar z)));\quad     v_2(\vec x)=\frac{|\omega|^2}{e}(1-\Phi^{\prime}(X(z,\bar z)));\label{005-3}\\ \Phi^{\prime\prime}(X)&-&(\Phi^{\prime}(X))^2+1=\frac{k}{|\omega|^2};\quad e\equiv \frac{r_2(\bar z)\bar\omega}{k_2(n-ia)}, \nonumber
\ea
and parameters must provide that the constant $e$ is real. Depending on the sign of $(\frac{k}{|\omega |^2}-1),$ one of two solutions is realized:
\be 
\Phi_1^{\prime}(X)=-\eta \cos (\eta X+\mu);\quad \Phi_2^{\prime}(X)=-\tilde\eta \ln(\cosh (\tilde\eta X+\tilde\mu));\quad (\frac{k}{|\omega |^2}-1)\equiv \eta^2>0; \,\, (\frac{k}{|\omega |^2}-1)\equiv -\tilde\eta^2<0.  \label{006-0}
\ee

For the second option, Eq.(\ref{003-3}) gives that $r_2(\bar z)=\gamma \bar z,$ constant $\frac{i\gamma }{n-ia}$ must be real, and:
\be 
v_1(\vec x) = \frac{4\lambda^2}{\beta}(1+\Phi^{\prime}(X));\quad v_2(\vec x)=\frac{\beta}{z\bar z}(1-\Phi^{\prime}(X));\quad \beta\equiv \frac{ik_2(n-ia)}{\gamma}.\label{006-1}
\ee 
Here, $\alpha (z)=i\lambda\ln (z),\,\, \bar\alpha (\bar z)=-i\lambda\ln (\bar z),$ and equation for $\Phi (X)$ takes the form:
\be 
\Phi^{\prime\prime}(X)-(\Phi^{\prime}(X))^2+1=k\lambda^2e^{\frac{X}{\lambda}}
\ee
See \cite{polyanin} about solutions of this equation in terms of special functions.

\section{Wave functions and electromagnetic fields.}

In the previous sections, we performed two consecutive transformations of Dirac operator with matrix potential using the first order intertwining operators similar to the SUSY intertwining in ordinary Quantum Mechanics. In that context, such operation is known as a second order reducible (i.e. factorizable) SUSY transformation 
\cite{AI}. Unlike that case, for the present problem with Dirac operator, the asymmetrical form of intertwining \cite{INP}, \cite{IN2022} was used in both steps. 

The resulting Dirac equation with potential which is a diagonal matrix with constant elements at the diagonal (like $U_2$ in (\ref{01-0})) is amenable to a simple analytic solution. Indeed, one of two components of $\Psi^{(2)}(\vec x)\equiv (\Psi_A^{(2)}(\vec x),\,\Psi_B^{(2)}(\vec x))^T$ can be excluded leading to the second order equation (Helmholtz equation) for another component:
\be 
(\Delta + m_1m_2)\Psi_A^{(2)}(\vec x)=0;\quad \Psi_B^{(2)}(\vec x)=\frac{2i}{m_2}\bar\partial\Psi_A^{(2)}(\vec x).
\label{A-1}
\ee
After separation of variables in (\ref{A-1}), its solution can be written as a linear combination with arbitrary complex coefficients $\sigma_{k_1k_2}:$
\be 
\Psi_A^{(2)}(\vec x)=\sum_{k_1,k_2} \sigma_{k_1k_2} \exp(k_1x_1) \exp(k_2x_2) \label{A-2}
\ee
where sum (actually, integral) is over $k_1,\,k_2$ - arbitrary complex constants such that $k_1^2+k_2^2=-m_1m_2.$ Coefficients in the sum have to be determined by boundary conditions for the wave functions.

According to the intertwining relations of the form (\ref{int}), the solutions $\Psi^{(1)}(\vec x)\equiv (\Psi_A^{(1)}(\vec x),\,\Psi_B^{(1)}(\vec x))^T$ of initial Dirac equation with potential $V_1(\vec x)$ can be constructed by sequential action of two intertwining operators $N_1\tilde N_1.$ The first operator $N_1$ intertwines two Dirac operators: initial one with potential $V_1$ and that with diagonal potential (\ref{V_2}) (see (\ref{int})). The second $\tilde N_1$ intertwines analogously Dirac operators with potential $V_2(\vec x)\equiv U_1(\vec x)\equiv W_1(\vec x)$ and operator with potential either $U_2(\vec x)$ or $W_2(\vec x),$ depending on exploring Subsection 4A,B or Subsection 4C. These intertwining operators have the general form (\ref{NM}), and the explicit expressions for the coefficients $A_k$ and $A(\vec x)$ were derived in Sections 3 and 4, correspondingly.

The two-component wave functions $\Psi^{(1)}(\vec x)$ obtained by given above procedure describe graphene and similar materials in external field: two-dimensional electrostatic plus non-homogeneous orthogonal magnetic. Analytical expressions for the strength of these fields are known from the analytical expression for initial potential $V_1(\vec x).$ The strength of electrostatic field is directed along the $(x_1, x_2)$ plane (see (\ref{A_0})):
\be 
\vec E(\vec x)=-\vec\nabla A_0(\vec x)=-\vec\nabla (f^2(\vec x)v_1(\vec x)), \label{A-3}
\ee
and the strength $B_3$ of magnetic one is (see (\ref{13-6})):
\be 
B_3(\vec x)= \triangle \ln (f(\vec x)).
\label{A-4}
\ee
Functions $f(\vec x)$ are different for different cases in Subsections 4A,B,C, and they can be calculated from the components $v_1(\vec x),\, v_2(\vec x)$ according to (\ref{13-2}). These components are given by (\ref{08-12}) and (\ref{08-22}) for {\bf 4A}; by (\ref{014-3}) for {\bf 4B}; (\ref{005-3}) and by (\ref{006-1}) for {\bf 4C}. The explicit expressions for these functions mainly in terms of trigonometric and hyperbolic functions lead to corresponding expressions for electromagnetic strengths in terms of the same elementary functions. 

For example of possible configuration of external fields we use the particular case of polynomial solution (\ref{014-6}) and (\ref{016-3}) for $c_1=c_2=0.$ The components $v_1(\vec x),\, v_2(\vec x)$ are given by (\ref{018-1}), and function $f(\vec x)$ is defined from:
\be 
f^4(\vec x)=\frac{v_2(\vec x)}{v_1(\vec x)}=\frac{2C}{L_1}[2a_1x_1^2-(2a_1+CL_1)x_2^2]^{-1}.
\nonumber 
\ee
According to (\ref{A-3}) and (\ref{A-4}), the strengths are:
\ba  
E_1(\vec x)&=&4\sqrt{2CL_1}a_1x_1\frac{2a_1x_1^2-3(2a_1+CL_1)x_2^2}{[2a_1x_1^2+(2a_1+CL_1)x_2^2]^{3}[2a_1x_1^2-(2a_1+CL_1)x_2^2]^{1/2}}; \nonumber\\ 
E_2(\vec x)&=&2\sqrt{2CL_1}(2a_1+CL_1)x_2\frac{6a_1x_1^2-(2a_1+CL_1)x_2^2}{[2a_1x_1^2+(2a_1+CL_1)x_2^2]^{3}[2a_1x_1^2-(2a_1+CL_1)x_2^2]^{1/2}}; \nonumber
\nonumber 
\ea 
and
\be 
B_3(\vec x)=2(4a_1+CL_1)(2a_1x_1^2+(2a_1+CL_1)x_2^2)(2a_1x_1^2-(2a_1+CL_1)x_2^2)^{-2}
\nonumber 
\ee 
with a possibility to choose arbitrary suitable values for all constant parameters.   

\section{Conclusions.}

The modification of well known method of SUSY Quantum Mechanics - asymmetrical intertwining relations - was used to build the massless two-dimensional Dirac equation with nontrivial matrix potential whose solutions can be found analytically. It was necessary to use factorizable second order intertwining which includes two steps: the first step allows to diagonalize the Dirac operator, and the second one - to connect the latter operator with explicitly solvable Dirac problem containing the diagonal matrix potential with constant elements. \\
{\bf Author Contributions.}

Conceptualization, writing and editing - M.V.I. and D.N.N. Both
authors have read and agreed to the published version of the
manuscript.\\
{\bf Funding.}

This research received no external funding.\\
{\bf Data Availability Statement.}

Data is contained within the article.\\
{\bf Conflicts of Interest.}

The authors declare no conflicts of interest.

\end{document}